\documentstyle[preprint,aps]{revtex}
\begin{document}
\def\be{\begin{equation}}
\def\ee{\end{equation}}
\title{Comparison of direct and Fourier space techniques in time-dependent
density functional theory}
\author{G.F. Bertsch$^{(a)}$\footnote{E:mail: bertsch@phys.washington.edu},
        Angel Rubio$^{(b)}$\footnote{E:mail: arubio@mileto.fam.cie.uva.es },
        and K. Yabana$^{(c)}$\footnote{E:mail: yabana@nucl.ph.tsukuba.ac.jp}}
\address{
$^{(a)}$Department of Physics and National Institute for Nuclear Theory,\\
University of Washington \\ Seattle, WA 98195\\
$^{b)}$ Departamento de F\'{\i}sica Te\'orica, Universidad de Valladolid,
E-47011 Valladolid, Spain\\ and Donostia International Physics
Center, San Sebastian, Spain\\
$^{(c)}$Institute of Physics, University of Tsukuba,\\
        Tsukuba 305-8571, Japan
}

\maketitle

\begin{abstract}

Several techniques have appeared in the literature to solve
the equations of time-dependent density functional theory.  We
compare the efficiency of different methods based on mesh
representations of the wave functions (direct and Fourier
space), taking as a test case
the calculation of the surface plasmon in the cluster Na$_8$.  For
smaller systems, the methods have comparable efficiency.  For
large systems the direct time method has a decided advantage in
computer storage requirements.  It is also more economical on arithmetic
operations, but is not as suited for parallel computing as the methods
based on a frequency representation.
\end{abstract}
\noindent
PACS: 31.10.+z,31.15.Ew,31.70.Hq,71.15.-m\\keywords: linear response,
density functional theory,electronic structure, electronic excitations,
time-dependent methods

\section{Introduction}
   The time-dependent local density approximation has proven to
be a useful tool to calculate the optical properties of finite
systems such as atoms, molecules, and atomic clusters
\cite{pe96,gr96,ru96,ya96,ya99,ru97,ca98,va99}. The
basic equation to be solved is conceptually very simple, little
more than the time-dependent Schr\"odinger equation for a
particle in a time-varying external field.  Many numerical methods
are in use to solve the equations.  On the one side there are
quantum chemistry methods based on atomic orbital representation
for the wave function, and on another side there are methods based
on mesh representations.  We only consider the latter here, but
even in this category there are a number of published techniques.
Most fundamentally, the time evolution can be calculated directly
or in Fourier space, i.e. in terms of frequencies.  The former
method is practically a necessity for dealing with very strong
external fields\cite{ul97,na99} and has been applied by two of us
(K.Y. and G.B.)
for the weak-field response as well\cite{ya96}. We shall call this
approach the ``nuclear physics"(NP) method, since the algorithms were
originally developed in that field for describing nuclear
reactions\cite{fl78}. The other methods we will consider\cite{ru96,va99}
solve equations in frequency space.  The method described in ref.
\cite{ru96} had its origins in condensed matter theory and uses
Fourier representation for both space and time; we shall call this
the ``condensed matter" (CMP) method. We also comment on ref. \cite{va99}
which uses Fourier space for the time but a real space mesh for
the spatial dependence\cite{comment_AO}.  Here the problem is 
cast into a matrix diagonalization in the particle-hole representation;
we shall call it the diagonalization method.

   In this work we will compare the CMP code and the NP
code for a specific system and present arguments for the scaling
properties of the respective algorithms for larger systems.  The
system we choose to study is the atomic cluster Na$_8$, and in
particular the surface plasmon excitation which is seen as a
strong peak at 2.5 eV excitation.  The TDLDA is not an exact theory
and it predicts a excitation energy at about 2.7 eV.  We shall
demand of both methods that they achieve within 0.1 eV of the
converged value.  It makes little sense to calculate to higher
precision in view of the intrinsic limitations of the theory.

   We shall now describe the various methods from a computational point
of view.  We shall use the symbol $N$ with subscripts for quantities that 
scale roughly as the size of the physical system under study, and $M$ for 
quantities that may be large but are independent of the size of the system. 
Important quantities common to the two codes are the number of
electrons $N_e$ and the number of mesh points, $N_G$ and $N_R$ for
real space and reciprocal space, respectively.
Additional quantities that play a
role are the number of frequencies to be calculated  $M_\omega$, and 
the number
of time steps to evolute the wave function in the real-time method, $
M_T$.  Also, in methods that rely on sparse matrix multiplication, we need
the number of nonzero entries in a row of the Hamiltonian, $M_H$, and
in iterative methods to solve large matrix equations we need the number
of iterations for convergence, $M_{it}$.  Finally,
the response function method usually requires a sum over unoccupied
states, $N_c$.  This notation is summarized in Table I.

We will use same energy functional for all methods, so the choice
of specific functional is not an issue in comparing the
methods.  As is commonly done, we calculate only the dynamics of
the valence electrons.  The core electrons are frozen and their
presence is treated by using a pseudopotential to describe the
ionic potential.  We use the pseudopotential construction of
Troullier and Martins \cite{tr91}, taking the nonlocal part by the
method of Kleinman and Bylander \cite{kl82} and including partial
core corrections for the exchange-correlation energy\cite{lo82}.
In this method, the local
pseudopotential is fixed to the value in a particular angular
momentum channel, and a nonlocal correction is made for other
channels.  Here we use the $l=1$ potential as the local potential,
and apply the nonlocal correction to the $l=0$ and
$l=2$ channels. The electron-electron interaction is taken in
the simple local-density approximation (LDA) given by Perdew and
Zunger \cite{pe81}.  More complicated functionals have better
predictive power for ground state properties \cite{ru97,ca98}, but
give only small improvement to the optical response of neutral
molecules. The proper description of the 1/r asymptotic behavior
of the potential is going to be very important to describe charged
systems, however for the aim of the present work this LDA
deficiency is not relevant.

The geometry of the Na$_8$ cluster was computed in ref.\cite{ru97},
and the lowest
energy structure found to be the bicapped octahedron (D$_{2d}$ symmetry). 
We use this structure in our comparison here.  It has an average Na-Na bond 
length of 3.38 \AA\ and a slight
deviation from the spherical symmetry.  This leads to a polarizability 
tensor with two different components and two close-lying peaks are obtained
in the photoabsorption cross section. 

\section{Theoretical methods}

Before describing in detail each of the two methods for representing
the wave functions (direct and Fourier space), we need to
comment on the choice of the spatial cell size and mesh size as
well as the time/frequency parameters (all are summarized in Table
I).

Since the wave functions are sensitive to boundaries, the
calculations must be made in a volume several Angstroms larger
than the size of the molecule or cluster.  Using both methods, we
determined how large a volume is needed to achieved 0.1 eV
accuracy on the various excitation energies of interest in the
system. We found that this is achieved in a spherical volume of
radius $R=8$ \AA\ using the NP code, and in a simple cubic
supercell of side 12.7 \AA\  using the CMP code. These have nearly
the same volume, and thus the same average distance from the
cluster to the boundary. We have checked the convergence of the
results by increasing the volume to a sphere of 12 \AA\ radius. The
value of the plasma frequency is reduced by a maximum of 0.1 eV,
that is, within the required accuracy.

We have used an uniform spatial grid with $\Delta x = 0.5$ \AA\ spacing. This
corresponds to a plane-wave cutoff energy of 6 Hartrees  in the Fourier
space method (see below).
Within this parameters, a stable time-step to perform the
time-evolution in the NP method is 
$\Delta t =0.003 \hbar/eV << \hbar (\Delta x)^2/m$. The required 0.1eV
accuracy in energy is obtained for total simulation times of 10
$\hbar$/eV. Similarly in the Fourier space method
we have taken a uniform frequency grid of
$M_{\omega}=100$ between 0 and 5 eV. Note that if the response is required for
larger frequencies we need to increase the number of points. The whole
response is obtained at once in the time evolution method (unless up to
energies of the order of $(\Delta t)^{-1}$). This is a great advantage
when the whole response is needed.

\subsection{ NP method }

This method uses a direct solution of the time-dependent single-electron Schr\"odinger
equation, 
\begin{equation}
\label{tdks} i \hbar \frac{\partial \phi_i({\bf r},t)}{\partial t} =
H_{KS}(t)\phi_i({\bf r},t) \,\,\,\, \,\,\,\, {(\rm i=1 \dots occ.)}
\end{equation}
where $H_{KS}$ is the Kohn-Sham Hamiltonian operator  
\be
H_{KS}(t)=- \frac{\hbar^2}{2m} \nabla^2 + V_{ion}({\bf r}) 
+ e^2\int d^3 r' \frac{ n({\bf r}',t)}{\vert {\bf r} -{\bf r}' \vert}
+ V_{xc}({\bf r},t)
\ee
and $n$ is the time-dependent electron density $n({\bf
r},t)=\sum_{i=1}^{occ}\phi_i^*({\bf r},t)\phi_i({\bf r},t)$.
In the
solution of this equation in the spatial and time variables
following the algorithm of ref.\cite{ya96}, there are two
time-consuming operations. One is multiplying the single-electron
Hamiltonian operator by the vector representing the wave function.
The dimensionality of the vector is the number of mesh points
$N_R$ times the numbers of electron orbitals $N_e$. The operator
is a sparse matrix with $M_H$ nonzero elements per row.  Thus the
basic operation requires about $ N_e N_R M_H $ complex floating point
operations.  The time evolution operator in the NP code is
implements by a power series expansion of the exponential operator
$\exp(-iH\Delta t)$ to fourth order.  A predictor-corrector cycle
requires two such operations.  Thus the method requires 8
Hamiltonian multiplications per time step.  Thus for $M_T$ time
steps the total number of floating point operations is given by

$$NP FPO : 10 N_e N_R M_H M_T$$

The sparseness of the Hamiltonian matrix in a real space
formulation is determined by the finite difference formula for
kinetic energy (nine-point formula in our case); and by the
nonlocal-projection parts of the potential. In total we have a number of
non-zero elements of the each Hamiltonian row 
$M_H\approx 100$ for the grid parameters used for Na$_8$.


The other time-consuming part of the NP algorithm is solving the
Poisson equation, which must be done twice at each time step.  The
NP code uses a multipole expansion combined with a relaxation
method to deal with the higher multipoles.  It is hard to estimate
the scaling properties of this part, but in the present study this
part of the computation takes 1.5 times as many operations as the Hamiltonian
multiplication operation. We shall assume the same factor for
estimating the scaling properties of the algorithm.  In principle,
the Poisson equation can be solved by methods that are of order
$N_R$ or $N_R \log N_R$, as multigrid or fast-Fourier
transformation, so this part should not dominate for large system.

Storage requirements are small: the vector wave function plus
$V_{Hartree}$ and $V_{ion}$ local potentials in Hamiltonian,
charge densities and some intermediate arrays. $V_{Hartree}$
requires a slightly larger volume because of the way the Poisson
equation is solved.

$$    {\rm NP\,\, storage:\,\,\,\,\,\,}N_R ( N_e+ 4.5) $$

This NP method is ideal to be combined with molecular dynamics
simulations for the ions because it uses only ground-state
occupied information and would scale roughly linearly
with the number of atoms in the system. There is not so much
book-keeping as in the usual perturbative formalism (no need for
storing the large set of unoccupied wave-functions and the large
dielectric matrices).

\subsection{CMP method}

Here the basic object of the calculation is the linear response to
an external field of some frequency $\omega$.  The linear response
matrix $\chi$ is constructed in momentum space with the following
matrix inversion
\be
\chi = ( 1- \chi_0 K)^{-1} \chi_0
\ee
where the independent particle response $\chi_0$ and the interaction
$K$ are matrices defined as follows.  The $\chi_0$ has elements
$G,G'$ given by \cite{hy87}
\be
\label{chi0}
\chi_0({\bf G},{\bf G}',\omega) = \frac{1}{\Omega} \sum_{kj} (f_k -f_j)
\frac{\langle k | e^{-i{\bf G}\cdot {\bf r}}| i \rangle
\langle i | e^{i{\bf G}'\cdot {\bf r}}| k \rangle} 
{\omega - \epsilon_j+\epsilon_i + i\eta}
\end{equation}
where $\Omega$ denotes the unit-cell volume,
$i,k$ label Kohn-Sham eigenfunctions
and $\epsilon_k$ and $f_k$ are the corresponding
eigenenergies and occupancy factors.  The sum goes over
$N_e$ occupied orbitals and $N_c$ empty orbitals.
The interaction $K$ is the
Fourier transform of the electron-electron interaction in the Kohn-Sham
equation, which is given in coordinate space by
\be
\label{K}
K({\bf r},{\bf r}')
= \frac{e^2}{\vert {\bf r}- {\bf r}' \vert} + \frac{\delta
V_{xc}({\bf r})}{\delta n({\bf r}')}
\ee
 
We now describe the computation starting with the Kohn-Sham wave functions 
and energies in a momentum space representation.  To evaluate
the independent particle response $\chi_0$ in eq.( \ref{chi0}), one first
calculates the particle-hole
matrix elements of the momentum operator and stores them in a table 
(or in disk).
This computational effort is of the order of
$N_e N_c N_G^2$ operations, and the table
size to be stored is $N_e N_c N_G$ complex numbers.     
Then the evaluation of eq. (4) requires $N_G^2$
matrix elements to be calculated, each requiring particle-hole
summation, to give 
$  \approx  2 N_G^2 N_e N_c $
operations for each frequency.
If one were to make
full space calculation, the number of empty orbitals summed in eq. (4) 
would be of
the same order as the dimensionality of the space. However, the
number of empty orbitals can be severely truncated without
effecting the long-wavelength dipole response. In the example, we
find $N_c=320$ is adequate, which is more than an order of
magnitude smaller than the size of the space and corresponds to
include unoccupied states up to 20 eV above the highest-occupied
orbital. This is  a reasonable approximation as we are
interested only in getting the optical spectra for excitation energies below 10
eV. This approximation is an important saving in building up 
the response matrix. 

One also truncates the calculation of the response matrix in another
way.  We have also assumed that the
off-diagonal elements of the response function are zero for
G-vectors outside an sphere of 1.25 \AA\ (that is to consider
$\approx$ 3200 points in the G-space). This corresponds to reducing the number
of matrix elements to be computed and stored to $N_G(N_G+4)/18$.
Note that the necessity to store the $N_G^2$ matrix puts a higher
demand on the computer memory than the NP method. The memory
required to store the $N_G^2$ complex, double-precision numbers in
the example problem is $164$Mb.

There are now three steps to evaluate eq. (3), two matrix 
multiplications
and a matrix inversion.   The matrices are not sparse, so the
matrix multiplications each cost $2 (N_G/3)^3$ arithmetical operations
\footnote{A small technical point should be mentioned, associated with
the divergence of the Coulomb
interaction at ${\bf G=G'}$.  This is dealt with\cite{hy87} by 
taking a numerical limit as $|{\bf G-G'}| \rightarrow 0$, and this adds
about 10\% to the number of operations for computing the matrix product.
}.
The matrix inversion is of the same order, requiring $(N_G/3)^3$ 
operations.  The total is
$ \approx 5 (N_G/3)^3.  $   
These represent the most computationally demanding steps in the CMP method,
given the truncation in the $N_c$. The computed $\chi$ is next transformed
to the coordinate space representation.  Using the fast Fourier transform,
this takes $\approx N_G^2 \log N_G$ operations. The  dynamical polarizability
can be now computed from $\alpha(\omega)=V_{ext}\chi V_{ext}$
as a matrix times vector multiplication. From this
one can easily extract the photoabsorption cross section 
$\sigma(\omega) = \frac{4\pi \omega}{c} Im \alpha(\omega)$.

Then the total computational effort in the CM method is:
 $${\rm CM \,\,FPO:\,\,\,\,\,\,\,} M_\omega ( N_c N_e (N_G/3)^2 + 5(N_G/3)^3)$$ 
with the last term dominant.  The  storage requirements
  for all the occupied and unoccupied wave
functions plus the whole complex response matrix
is
$${\rm CM \,\, storage:\,\,\,\,\,} (N_e+N_c)N_G + 2(N_G + N_G^2/9) + N_cN_eN_G/3$$

To achieve the targeted energy convergence with this algorithm,
the momentum space mesh was chosen to correspond to a
simple cubic supercell of $L= 12.7$
\AA\ on a side.  This implies that the mesh spacing in momentum space is
$\delta k = 2 \pi /L = 0.137 $ \AA. The momentum space
representation takes all the points within a sphere of radius
$k_{max} = 1.83$ \AA\ (that corresponds to a plane-wave cutoff
energy of 12 Ry). The size of the vector in the momentum
representation is thus $ N_G = 4 \pi (k_{max}/\Delta k)^3/3
\approx 10,000$. Note that this is slightly smaller than the
number required for the coordinate space representation, however
we need to stress that a larger number of G-vectors are needed to
describe the action of the potential on a wave-function ($V\psi$
corresponds to a convolution in Fourier space).  Finally, an additional
numerical parameter is the imaginary part of the frequency $\eta$,
which we have taken as $\eta=0.05$ eV to produce a
resolution of 0.1 eV in the spectral features.

In  the discussion below we have not include the computational
requirements to perform the ground state calculations, occupied
and unoccupied orbitals.  This could be a major storage
bottle-neck for very large systems as the calculation of a large
set of unoccupied wave functions  has a cubic scaling of the number
of atoms in memory and computing time. In the present calculation
this initialization process takes 10\% of the total computational time.

\subsection{Other methods}

We mention here two other methods from a computational point of
view. Since we have not carried out numerically computations on
our test problem with these methods, the discussion will be brief.

\subsubsection{Modified Sternheimer method}
   The modified Sternheimer method was first applied to the time-dependent
Kohn-Sham equation for atomic excitations\cite{st95}, and has
since been applied to the dielectric response of crystals using
the momentum space representation \cite{dc96} and to the finite
system C$_{60}$ \cite{iw00} using the coordinate space representation.
Here one solves an inhomogeneous equation for the
perturbed wave functions $\phi_i^{\pm}$ using an iterative method. 
The perturbation is
a sinusoidal potential field combining the external field $V_{ext}$
and the internal field from the time-varying electron density.  The
equations are
\be
(\epsilon_I - H^0_{KS} \pm \omega +i \eta) \phi_i^{\pm} 
= \hat P V_i\label{sternheimer}
\ee
where
$$
V_i = (V_{ext} + K \delta n) \phi_i
$$
and 
\be
\delta n = {\rm Re} \sum_i \phi_i ( \phi_i^+ + \phi_i^-).
\ee
$\hat P$ is a projection operator removing occupied orbitals.
In ref. \cite{iw00}, the two equations are constructed in coordinate
space and solved with a double iteration.
One makes a guess for the density $\delta n$, and solves eq.
(\ref{sternheimer}) by the conjugate gradient method.  $\delta n$
is refined from the resulting $\phi_i^{\pm}$ again with the conjugate
gradient method, and the process is
repeated to convergence.  The numerical cost will thus depend largely
on the cost of the Hamiltonian operation which is $\approx M_H N_R N_e$ in coordinate
space,
and the number of iterations $M_{it}$ required to get a converged solution.
Remembering also that
frequency space methods need $M_\omega$,
the number of frequencies to be examined, 
the computational cost of this method is 
\be
{\rm Modified\,\, Sternheimer\,\,(real\,\, space):\,\,\,\,\,\,} M_\omega M_{it}M_H N_R N_e
\ee
The method can be used in this form for nonresonant frequencies, but
near the eigenfrequencies the nearby singularities in eq. (6) must be removed
for the conjugate gradient method to converge.  
Thus this method would be similar to methods utilizing the particle-hole 
representation in needing a considerable number of the wave functions 
and eigenenergies of unoccupied states.  
The singularities are removed by projecting on the
unoccupied wave function subspace the right hand side of eq. (6),
$$
V_i' =V_i - \sum_j \phi_i (\phi_i,V_i).
$$
The desired wave functions $\phi_i^{\pm}$ are obtained from the projected
solutions $\phi_i'^{\pm}$ by
$$
\phi_i^{\pm} = \phi_i'^{\pm} + \sum_j { \phi_i (\phi_i,V_i) \over
\epsilon_j-\epsilon_i -\omega -i \eta}.
$$
It is difficult to give an {\it a priori} estimate of $M_{it}$ or
its size-scaling properties (although with our notation we have 
assumed that it does not grow with $N$).  Unfortunately, our implementation
of eq. (7) still left the convergence somewhat erratic.
Typically it takes of the order of
$M_{it}\approx 1000$ iterations of the double loop to get convergence. 
Thus it would require some improvement of the algorithm to make it
attractive to apply to large systems.

The momentum space implementation of the modified Sternheimer method
is similar. This method also needs the conditioning step 
for convergence of the CG iteration.
The main difference is in the Hamiltonian multiplication, which   
here requires $\approx 2 (N_G/3)^3$ operations as discussed 
in Sect. IIB.  Thus
the total is
\be
{\rm Modified\,\, Sternheimer\,\, (momentum\,\,  space):\,\,\,\,} 
2 M_\omega M_{it}(N_G/3)^3
\ee  
Because the Hamiltonian operation is more costly in momentum space,
this method is
probably not competitive to the others, unless it were the case 
that the convergence of the iteration were intrinsically much more 
reliable.

\subsubsection{Diagonalization method}  
     The frequency-space methods discussed so far have relied
in some way on operator inversion.  It is also possible to 
cast the problem as one of matrix diagonalization.  This
method was applied to cluster excitations in the TDLDA 
by Vasiliev et al.
\cite{va99}.  The authors start from a basis in coordinate
space and construct Kohn-Sham orbitals for both occupied and
empty states as is done in the CMP method, but representing the
orbitals in coordinate space mesh, as in the NP method.
The storage requirement for the
orbitals is $\approx (N_c + N_e) N_R$, which is larger than
in the NP method but smaller than in the CMP method. 

The next step of the calculation is to construct the matrix to
be diagonalized.  
The eigenvalue equation to be solved is
\be
{\bf R} {\bf F}_n = \omega_n^2 {\bf F}_n
\ee
where ${\bf F_n}$ are the eigenvectors and ${\bf R}$ is a matrix.  Its
elements are 
\be
R_{\alpha,\alpha'} = (\epsilon_i-\epsilon_j)^2 \delta_{\alpha,\alpha'}
+ 2 \sqrt{ (\epsilon_i-\epsilon_j)(\epsilon_i'-\epsilon_j')}
K_{\alpha,\alpha'}
\ee
where the indices $\alpha=(ij), \alpha=(i',j')$ label combinations of
unoccupied orbitals $i$ and occupied orbitals $j$.  The interaction
matrix elements $K_{\alpha,\alpha'}$ are simply the particle-hole
matrix elements of the residual interaction, eq.(\ref{K}).  
There is a substantial computational cost in construct the interaction
matrix $K$.  
A straightforward transformation from the
coordinate space to the particle-hole representation requires 
$\approx N_R^2 N_e^2 N_c^2$ operations for the Coulomb interaction.
However, this is reduced considerably by using an efficient method
to solve the Poisson equation\cite{private}.  For example, using the
fast Fourier transform one may find the Coulomb field for a given particle-hole state
taking only $N_R \log N_R$ operations.  Saving the Coulomb field in
the coordinate representation, the matrix element to a
given final state takes $\sim N_R$ operations.  The effort of solving
the Poisson equation is thus distributed over the number of final
states, and the operations to construct the full matrix
has a leading dependence $ N_R N_e^2 N_c^2$, the scaling appropriate
for the local part of the interaction\footnote{However, in the
implementation of ref. \cite{va99}, the Poisson solver in fact is the
most costly operation.}. Once the
matrix is constructed, the diagonalization requires $\approx (N_cN_e)^3$
operations.  However, taking the $N$ values from Table \ref{symbols}, the
matrix diagonalization effort is small compared to that needed to
construct the matrix.  We have therefore taken that step
to assign this method's size scaling in Table II.

\section{Numerical results}

We will discuss in detail the physical quantities computed in the NP and 
CMP methods and refer to \cite{va99} for the results using the
diagonalization method. We want
to stress that the three approaches must give the same values
if the numerical parameters are chosen with fine enough grids and
large enough cutoffs to get converged results.

With the parameter sets chosen for the two methods, the results
are quite similar.  In Table \ref{plasmon} we show calculated Kohn-Sham
energies and the surface plasmon energy.  The first entry
$\epsilon_1$ is the Kohn-Sham eigenvalue of the most bound
orbital.  The absolute energies have no significance in the
supercell method, because the absolute Coulomb potential is
undefined.  Therefore, for this entry we give the value from the
NP code and set the scale of the CMP energies at that value.  The
next three rows correspond to the other bound orbitals use the
$G=0$ point of the Brillioun zone for the CMP values.  We can see
that the methods agree to within less than 0.1 eV.  The next entry
is the lowest unoccupied orbital.  This is significantly different
for the two methods.  This orbital has sufficient extension to
have its energy sensitive to the boundary, which of course is
different for the two methods.  We confirm the boundary
sensitivity in the CMP code by calculating the energies at other
points in the Brillouin zone.  Differences are less than 0.1 eV
for occupied orbitals, but reach 0.2 eV for the lowest unoccupied
orbital. This last point indicates the fact that the empty orbitals
are more sensitive to the boundary conditions and in the periodic
supercell they feel the potential from the other clusters. 

We have also presented in Table \ref{plasmon} the results of the NP method 
\footnote{The plasmon frequency is sensitive to the core-exchange correction
at the level of 0.1 eV.  We have included that correction in 
$H_{KS}$ it improves the description of the 
structural properties of Na metal.  We note that the
result without core corrections (2.89 eV) it is very 
close to the jellium value (2.9 eV). 
}.
We have also checked the
convergence of the plasmon frequency with respect to the cell size and found
that this value is converged to less than 0.01eV for a sphere of R=12 \AA.
The fully converged value in the NP method is 2.65 eV. The difference with the
experimental value of 2.53 eV can be attributed to deficiencies in the
LDA approximation
as well as for finite temperature effects in the experiments\cite{ru97}.

In Table \ref{polarizability} we summarize the results for the static averaged 
electrical polarizability of Na$_8$ obtained by the different methods. 
The agreement among the different approaches is very good and the remaining
difference with experiments can be again assigned to  
core polarization, exchange-correlation and temperature effects. These
effects tends
to increase the polarizability bringing the computed values close to the
experiments \cite{ru97}.

\section{Conclusions}

In the theory of electronic excitations of finite many-electron systems,
the time-dependent Kohn-Sham equation with an adiabatic local density
approximation for the interaction energy function offers an attractive
compromise towards the goals of accuracy and computational practicality.
But even within the  TDLDA scheme there are several methods in use,
and our purpose was to compare them on the same footing by applying them
to the same physical problem, and demanding the same accuracy.  The goal 
is to gain a general understanding 
of the numerical resources (total numbers of arithmetic operations and
computer memory)  required by the different methods.  One can then 
extrapolate to large systems and make a judgment on which methods offer
the best prospects.

We have only considered methods based on a grid representation of the
electron wave functions, and have concentrated on two algorithms, 
the NP method in real time and real space, and the CMP method in
Fourier transformed time and space.

We chose to study the response of the Na$_8$ cluster around the
surface plasmon excitation energy.  The two methods turned out to
have comparable requirement on arithmetic operations.  However, it
should also be noted that the computational work increases with the
range of frequencies that one studies in the CMP method, but not in
the NP method.  With latter, the entire response is obtained from
a single calculation.

In comparing the two methods to ascertain their scaling with the
size of the system $N$, we have deliberately ignored the first task
in either method, the construction of the eigenstates of the
static Kohn-Sham operator.  In the NP method only the occupied orbitals
are needed, but in the CMP method one also needs a large number of unoccupied
orbitals as well.  Their calculation scales like $N_e^3$ in principle,
but in practice this phase of the computation is short compared to the
dynamic calculation and so we ignore it.  Let us now compare the
scalings by taking the expressions in Table II, dropping the subscripts
on the $N$ quantities.  The NP method thus scales as $N^2$.  This behavior
was also found studying the excitations of long carbon molecules\cite{ya99}.
The CMP method has a poorer scaling behavior, namely $N^3$.  We also
considered two other methods without however examining them in as much 
detail.  In principle, the modified Sternheimer method in coordinate
space can achieve $N^2$ scaling without the cost of the large $M_T$
factor of the real-time method.  However, we did not find a reliably
converging iteration procedure to solve the basic inhomogeneous linear
equation set.  The final method we discussed, the diagonalization method using
real space and Fourier
time, seems to have a poorer $N$-scaling than the others, but may be
advantageous in some circumstances (see below).

Besides arithmetic operations, storage can play a role in the 
practicality of the different algorithms for large systems.  Here we
find that 
the storage requirements are grossly different for the NP and CMP methods,
favoring the NP approach.  From Table II, it has a $N^2$ scaling while
the CMP method has an $N^3$ behavior.  This is already significant in
the Na$_8$ system we studied, as may be seen from Table III.  

Thus our results favor the real-time and real-space methods, offering
economy in both storage and arithmetic operations.  However, there
are a number of caveats.  We have not considered the suitability of
the different algorithms for parallel computing. In
a parallel computing environment, the frequency-space methods gain favor
because the $M_\omega$ factor can be trivially absorbed in the parallel
processing.  In addition, the diagonalization method can benefit from the parallel
computation of different rows of the matrix.  Also the sparseness of 
the Hamiltonian matrix is important for the real space method; this
would be lost if for example the energy functional used the full Fock
exchange interaction.

Finally, we mention two nonnumerical benefits of the real-time method:
as was said earlier, it 
is nonperturbative and therefore allows effects of large fields
to be calculated with the same effort.  
And it uses the same energy functional (permitting the program to
call the same subroutine) for the dynamic calculation as
for the static calculation to prepare the ground state.

\section{Acknowledgment}
We are grateful to J. Chelikowsky and I. Vasiliev for communications
and providing us with their computer code.
This work was supported by the Department of Energy under
Grant FG06-90ER-40561, by the DGES (PB98-0345) and
JCyL (VA28/99), and by  he Grant-in-Aid for Scientific Research from the
Ministry of Education, Science and Culture (Japan), No. 11640372.
  AR acknowledges the hospitality of the Institute for 
Nuclear Theory where this work was started and the computer time 
provided by the C$^4$ (Centre de Computaci\'o i Comunicacions de Catalunya).

\begin{table}
\caption{Symbol definitions for quantities pertaining to the computational
effort required by the various algorithms discussed in the main text,
and their values.}
\begin{tabular}{cc|cc} 
Symbol & Meaning & NP method & CMP method\\ \hline
 $M_T$ & time steps & $10^4$ &  -  \\
 $M_\omega$ & number of frequencies& - & $10$ \\
 $M_H$ & nonzero elements in H matrix row & 100  &  - \\
 $M_{it}$ & iterations in conjugate gradient method & - & - \\
 $N_R$ & real-space points & $17,000$ & - \\
 $N_G$ & reciprocal-space points & - & 9,771\\
 $N_e$ & number of electron orbitals (occupied states)& 4 & 4\\
 $N_c$  & unoccupied states & - & 320  \\
\end{tabular}
\label{symbols}
\end{table}

\begin{table}
\caption{Leading-order for the size scaling of various algorithms for
TDLDA--general comparison: floating point operations (FPO) and memory
requirements.}
\begin{tabular}{l|ll}
Method         &   FPO    & Memory   \\
\hline
 NP &   $N_e N_R M_H M_T$  &  $N_R(N_e+4.5)$\\
CMP  &  $ 5 M_\omega (N_G/3)^3 $& $N_G^3/9$\\
Modified Sternheimer & $M_\omega M_{it}M_HN_eN_R$   & $N_R(N_e+N_c)$   \\
 Diagonalization  & $N_c^2 N_e^2 N_R$ &  $(N_c N_e)^2$\\
\end{tabular}
\label{scaling}
\end{table}

\begin{table}
\caption{Comparison of computational difficulty of NP and CMP
methods for Na$_8$}
\begin{tabular}{c|cc}
Resource  & NP & CMP\\ \hline Memory (MBy) & 7  & 350 \\
 Floating point operations & $1.5 \times 10^{12}$  & $1.7 \times 10^{12}$ \\
\end{tabular}
\label{resources}
\end{table}

\begin{table}
\caption{Orbital energies $\epsilon_i$ and surface plasmon energy
$\omega_{M}$ in Na$_8$. For comparison in parenthesis we show the
result of a calculation within the NP method without including
partial core corrections in the pseudopotential generation and
time evolution.}
\begin{tabular}{c|ccc}
Energy & NP  & CMP &Exp. (eV)\\ \hline
 $\epsilon_1$&  -4.63 
&-4.63&\\
 $\epsilon_2$ & -3.41 
&-3.35 & \\
 $\epsilon_3$ & -3.00 
&-2.97& \\
 $\epsilon_4$ & 3.00  
& -2.97&\\
 $\epsilon_5(LUMO)$ & -1.88 
&-2.01&\\
 $\omega_M$ &  2.77 
& 2.6 & 2.53 \cite{he87} \\
\end{tabular}
\label{plasmon}
\end{table}

\begin{table}
\caption{Static polarizability of Na$_8$ (\AA$^3$)}
\begin{tabular}{c|cccc}
Exp.  & NP & Atomic &  CMP &  All-electron\\ \hline 128.7
\cite{kn85} & 103  &  117\cite{va99}  & 119  & 114.9 \cite{ca99}\\
\end{tabular}
\label{polarizability}
\end{table}

\begin{references}
\bibitem{pe96} M. Petersilka, U. J. Gossmann and E. K. U. Gross,
Phys. Rev. Lett. {\bf 76}, 1212 (1996).
\bibitem{gr96} E.K.U. Gross, J.F. Dobson, and M. Petersilka, in
{\it Density Functional Theory II}, edited by R.F. Nalewajski,
``Topics in Current Chemistry", Vol 181 (Springer, Berlin, 1996)
p.81
\bibitem{ru96} A. Rubio, et al., Phys. Rev. Lett. {\bf 77} 247 (1996).
\bibitem{ya96} K. Yabana and G.F. Bertsch, Phys. Rev. B{\bf 54} (1996) 4484.
\bibitem{ya99} K. Yabana and G.F. Bertsch, Int. J. Quantum Chemistry {\bf 75} 
(1999) 55. 
\bibitem{ru97} A. Rubio, J.A. Alonso, X. Blase, and S.G. Louie,
Int. J. Mod. Phys. B {\bf 11}, 2727 (1997), and reference therein.
\bibitem{ca98} M.E. Casida, et al., J. Chem. Phys. {\bf 108} 4439 (1998).
\bibitem{va99} I. Vasiliev, S. Ogut, and J. Chelikowsky, Phys.
Rev. Lett. {\bf 82} 1919 (1999).
\bibitem{private}  I. Vasiliev, private communication.
\bibitem{ul97} C.A. Ullrich, P.-G. Reinhard and E. Suraud, J.
Phys. B: At. Mol. Opt. Phys. {\bf 30} 5043 (1997).
\bibitem{na99} R. Nagano, K. Yabana, T. Tazawa, Y. Abe, 
J. Phys. B: At. Mol. Opt. Phys. {\bf 32} L65 (1999).
\bibitem{fl78} H. Flocard, et al., Phys. Rev. C{\bf 17} 1682 (1978).

\bibitem{comment_AO} Different representations of the wave
function based on linear combination of atomic orbitals (AO) have
been used in the literature to address the linear and nonlinear
response of molecules\cite{ca98,sv99} with quite a good success.
As compare to grid or plane-wave-based representations, the main
advantage of the AO representation stems from the small number of
basis needed to expand the wave functions and Hamiltonian matrix
elements. On the other hand, the matrix elements in the response
function  cannot  be calculated as easily as in a plane-wave
reresentation and the check of convergence with respect to the
size of the AO-basis set is rather difficult (usually the number of AO's
is much less than $N_G$).

\bibitem{sv99} S.J.A. van Gisbergen, et al,
Phys. Rev. Lett. {\bf 83}, 694 (1999); Phys. Rev. Lett. {\bf 78}, 3097 (1997)
\bibitem{tr91} N. Troullier and J.L. Martins, Phys. Rev. B{\bf 43} 1993 (1991)
\bibitem{kl82} L. Kleinman and D. Bylander, Phys. Rev. Lett. {\bf 48} 1425 (1982).
\bibitem{lo82} S.G. Louie, S. Froyen, and M.L. Cohen, Phys. Rev. B {\bf 26},
 1738 (1982)
\bibitem{pe81} J. Perdew and A. Zunger, Phys. Rev. B {\bf 23} 5048 (1981).
\bibitem{hy87} M.S. Hybertsen and S.G. Louie, Phys. Rev. B {\bf 35}, 5585 (1987).
\bibitem{st95} M. Stener, P. Decleva and A. Lisini, J. Phys. B{\bf 28} 4973
(1995).
\bibitem{dc96} A. Dal Corso, F. Mauri and A. Rubio, Phys. Rev. B {\bf 53},
15638 (1996).
\bibitem{iw00} J.-I. Iwata, K. Yabana and G.F. Bertsch, Nonlinear Optics,
to be published.
\bibitem{he87} W.A. de Heer, et al, Phys. Rev. Lett. {\bf 59}, 1805 (1987);
C.R.C. Wang, et al, J. Chem. Phys, {\bf 93}, 3789 (1990);
 W.A. de Heer, Rev. Mod. Phys. {\bf 65}, 611 (1993).
\bibitem{kn85} W.D. Knight, K. Clemenger, W.A. de Heer, and W.A. Saunders,
Phys. Rev. B{\bf 31} 2539 (1985).
\bibitem{ca99} P. Calaminici and A. M. K\"oster, J. Chem. Phys. {\bf 111} 4613 (1999).


\end{references}
\end{document}